\documentclass{PoS}

\usepackage{url}

\graphicspath{{figures/}}

\title{Performance of a small size telescope (SST-1M) camera for gamma-ray astronomy with the Cherenkov Telescope Array}

\ShortTitle{Calibration of the SST-1M camera}
\author{
I.~Al~Samarai\textsuperscript{a},
\speaker{C.~Alispach}\textsuperscript{a},
F.~Cadoux\textsuperscript{a},
V.~Coco\textsuperscript{a},
D.~della Volpe\textsuperscript{a},
Y.~Favre\textsuperscript{a},
M.~Heller\textsuperscript{a},
T.~Montaruli\textsuperscript{a},
A.~Nagai\textsuperscript{a},
T.R.S.~Njoh~Ekoume\textsuperscript{a},
I.~Troyano Pujadas\textsuperscript{a},
E.~Lyard\textsuperscript{g}, 
A.~Neronov\textsuperscript{g},
R.~Walter\textsuperscript{g},
V.~Sliusar\textsuperscript{g},
E.~Mach\textsuperscript{h},
J.~Micha{\l}owski\textsuperscript{h},
J.~Niemiec\textsuperscript{h},
J.~Rafalski\textsuperscript{h},
K.~Skowron\textsuperscript{h},
M.~Stodulska\textsuperscript{h},
M.~Stodulski\textsuperscript{h},
T.~Bulik\textsuperscript{d},
M.~Grudzi{\'n}ska\textsuperscript{d}, 
M.~Jamrozy\textsuperscript{b},
M.~Ostrowski\textsuperscript{b},
{\L}.~Stawarz\textsuperscript{b},
A.~Zagda\'{n}ski\textsuperscript{b}, 
K.~Zi{\c e}tara\textsuperscript{b},
P.~Pa{\'s}ko\textsuperscript{i},
K.~Seweryn\textsuperscript{i},
J.~Borkowski\textsuperscript{c},
R.~Moderski\textsuperscript{c}, 
J.~Kasperek\textsuperscript{k},
P.~Rajda\textsuperscript{k},
D.~Mandat\textsuperscript{m}, 
M.~Pech\textsuperscript{m},
P.~Schovanek\textsuperscript{m},
P.~Travnicek\textsuperscript{m} for the CTA SST-1M Project. \\
\textsuperscript{a}\textit{DPNC - Universit\'e de Gen\`eve,  Switzerland} \\
\textsuperscript{b}\textit{Astronomical Observatory, Jagiellonian University, Poland} \\
\textsuperscript{c}\textit{Nicolaus Copernicus Astronomical Center, Poland} \\
\textsuperscript{d}\textit{Astronomical Observatory, University of Warsaw, Poland} \\
\textsuperscript{g}\textit{ISDC, Observatoire de Gen\`eve, Universit\'e de Gen\`eve, Switzerland} \\
\textsuperscript{h} \textit{Instytut Fizyki J{\c a}drowej im. H. Niewodnicza{\'n}skiego Polskiej Akademii Nauk, Poland} \\
\textsuperscript{i}\textit{Centrum Bada{\'n} Kosmicznych Polskiej Akademii Nauk, Poland} \\
\textsuperscript{k}\textit{AGH University of Science and Technology, Poland} \\
\textsuperscript{m}\textit{Institute of Physics of the Czech Academy of Sciences, Czech Republic.} \\
\textsuperscript{1}\textit{https://www.cta-observatory.org/} \\
E-mail: \email{imen.alsamarai@unige.ch}

}

\abstract{The foreseen implementations of the Small Size Telescopes (SST) in CTA will provide unique insights into the highest energy gamma rays offering fundamental means to discover and understand the sources populating the Galaxy and our local neighborhood. Aiming at such a goal, the SST-1M is one of the three different implementations that are being prototyped and tested for CTA. SST-1M is a Davies-Cotton single mirror telescope equipped with a unique camera technology based on SiPMs with demonstrated advantages over classical photomultipliers in terms of duty-cycle. 

In this contribution, we describe the telescope components, the camera, and the trigger and readout system. The results of the commissioning of the camera using a dedicated test setup are then presented. The performances of the camera first prototype in terms of expected trigger rates and trigger efficiencies for different night-sky background conditions are presented, and the camera response is compared to end-to-end simulations.  
}

\FullConference{35th International Cosmic Ray Conference --- ICRC2017\\
		10--20 July, 2017\\
		Bexco, Busan, Korea}

\begin{document}

\section*{Introduction}

The Cherenkov Telescope Array (CTA)\cite{CTAScience} will consist of three types of telescopes: large (LST), medium (MST) and small (SST) size telescopes. The SSTs are dedicated to the observation of gamma rays with energy between a few TeV and a few hundreds of TeV. The SST array is expected to have 70 telescopes of different designs.
The single-mirror small size telescope (SST-1M) is one of the proposed telescope designs under consideration for the SST array with a Davies-Cotton optics. It is equipped with a 4m-diameter segmented mirror dish as well as an innovative camera based on silicon photomultipliers (SiPMs).

\section{The SST-1M camera}

The SST-1M camera~\cite{camera2016} is composed of the Photo-Detection Plane (PDP) equipped with the SiPMs, and the fully digital readout and trigger system (DigiCam).

\subsection{The Photo-Detection Plane (PDP)} 
The PDP sensitive area has a hexagonal geometry composed of 1296 pixels grouped in 108 modules of 12 pixels each. The PDP pixels are formed by a  hexagonal hollow light-funnel coupled to a large area, hexagonal SiPM sensor from Hamamatsu. The 12-pixel modules are composed by a preamplifier board and a slow control board (SCB).  
The SCB serves to monitor or control the slow control parameters of each sensor (such as bias voltage and temperature) and to stabilize its operational point by actively adapting the over-voltage according to the temperature variations.

\subsection{The trigger and readout system (Digicam)}

DigiCam provides a continuous digitization of the signals received by the PDP using the same data to produce a trigger decision.
Analog signals from the PDP are transferred to the digitizer boards via standard CAT5 cables for digitization at a sampling rate of 250~MHz by 12-bit FADCs. The samples are then serialized and sent in packets through high speed multi-gigabit serial digital data interfaces to the Xilinx XC7VX415T FPGA, where they are pre-processed and stored. In order to reduce the size of the data processed by the trigger card, the digitized signals are grouped in sets of three adjacent pixels (called patches) and clipped at 8 bits. The trigger decisions are taken based on the signal level in specified geometrical patterns of patches, called clusters, in the lower resolution copy of the image. If an event is selected, the corresponding full resolution data stored in the digitizer boards are sent to the central acquisition system of the telescope by the master trigger card via a 10 Gbps fiber Ethernet link.

\section{Camera commissioning} 
In November 2016, SiPM sensors constituting the PDP have been mounted on the camera support. Digicam micro-crates equipped with the digitization and triggering boards have been installed, and the full camera hardware implementation has been validated. The triggering and readout software with a flow of real data could be tested for the first time. 

To emulate the light from the showers, which will impinge on the PDP, the camera test setup (CTS) is a LED board designed as a set of printed circuit boards covering one third of the camera in addition to a central zone common to the three sectors. The LED board houses two LEDs pointing to each pixel, one pulsed (AC mode) emulating the signal, and one in continuous light mode (DC mode) emulating the night-sky background (NSB). The LED board calibration procedure in AC and DC modes is presented in \cite{camera2016, cyrilICRC2017}. 

\subsection{First camera prototype measurements} 

\begin{description}
\item[Calibration parameters:]
During real operations in CTA of the SST-1M telescope, calibration runs will be performed prior to data taking. In particular, it is important to assess the dark count rate in the camera as it introduces a bias in the determination of calibration parameters. 
The mean dark count rate and the cross-talk probability per pixel are both extracted from the single photo-electron spectrum (SPE) measured in dark conditions by shielding the camera from external light (equivalent to an NSB level of $\sim$ 3 MHz/pixel).  
 The mean dark count rate is found at 2.05$\pm$0.2 MHz per pixel, while the optical cross-talk probability is 8\%. 
Multi-photoelectrons spectra (MPE) are acquired using the calibrated light emitted by the CTS. The gain, representing the conversion factor from the Least Significant Bit (LSB) to  the number of  PE can be extracted from the MPE. It is found at 5.6 on average, while the dispersion over one third of the camera is 3\%. In addition, the gain drop as a function of the NSB pointed out in \cite{matthieuSPIE, camera2016} has been measured and accurately parametrized on each pixel in the sector covered by the DC-mode LEDs of the CTS. 
The camera calibration and the related systematic errors are presented in detail in \cite{cyrilICRC2017}.
In addition to the camera calibration, such parameters serve as input to the camera simulations. 

\begin{figure}  
\centering
\begin{minipage}{0.45\textwidth}
  \includegraphics[width= \textwidth]{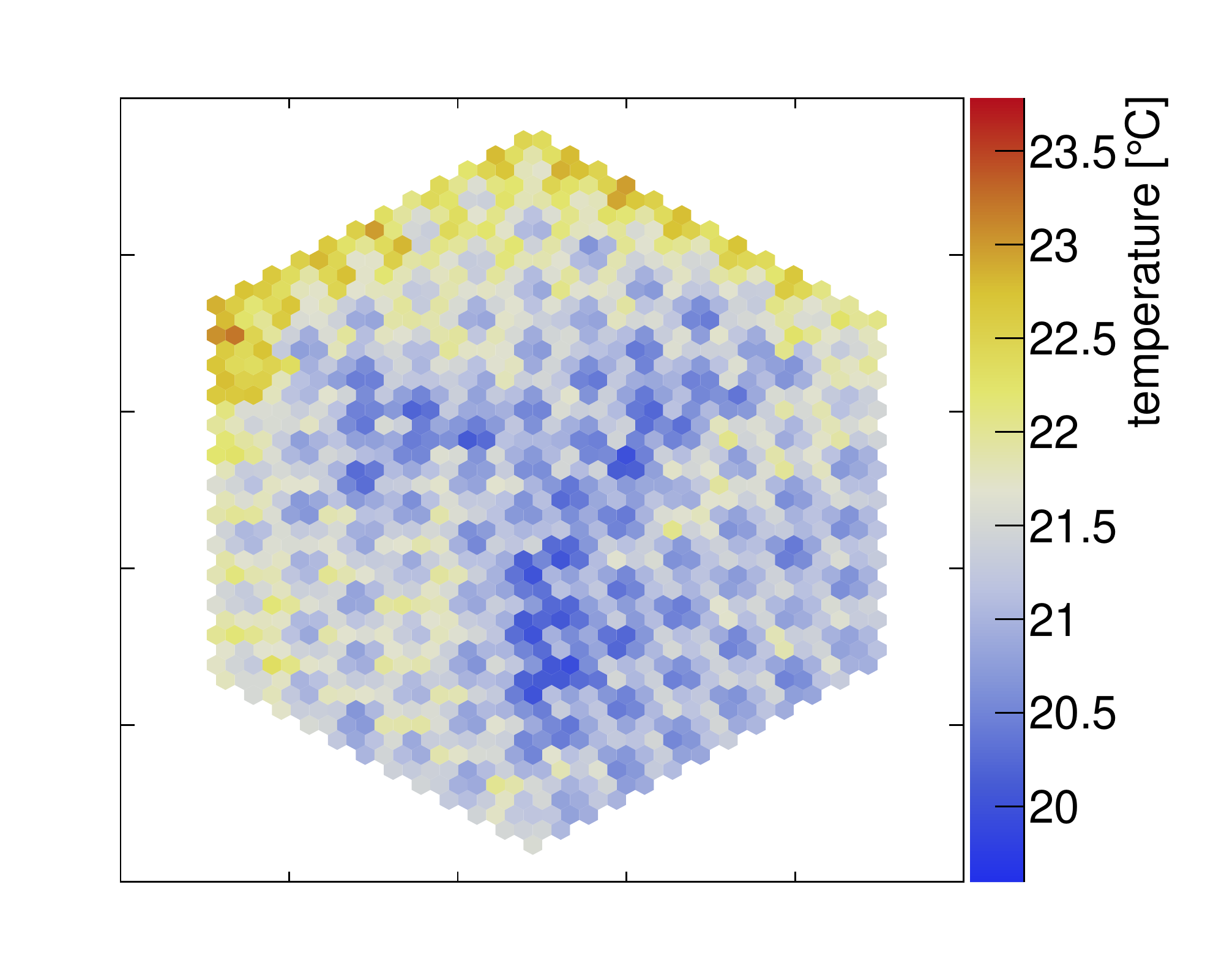}
 \caption{The temperature map obtained over the full camera. }
 \label{fig:mpegain}
\end{minipage}
\hspace{0.5 cm}
\begin{minipage}{.475\textwidth}
  \includegraphics[width= \textwidth]{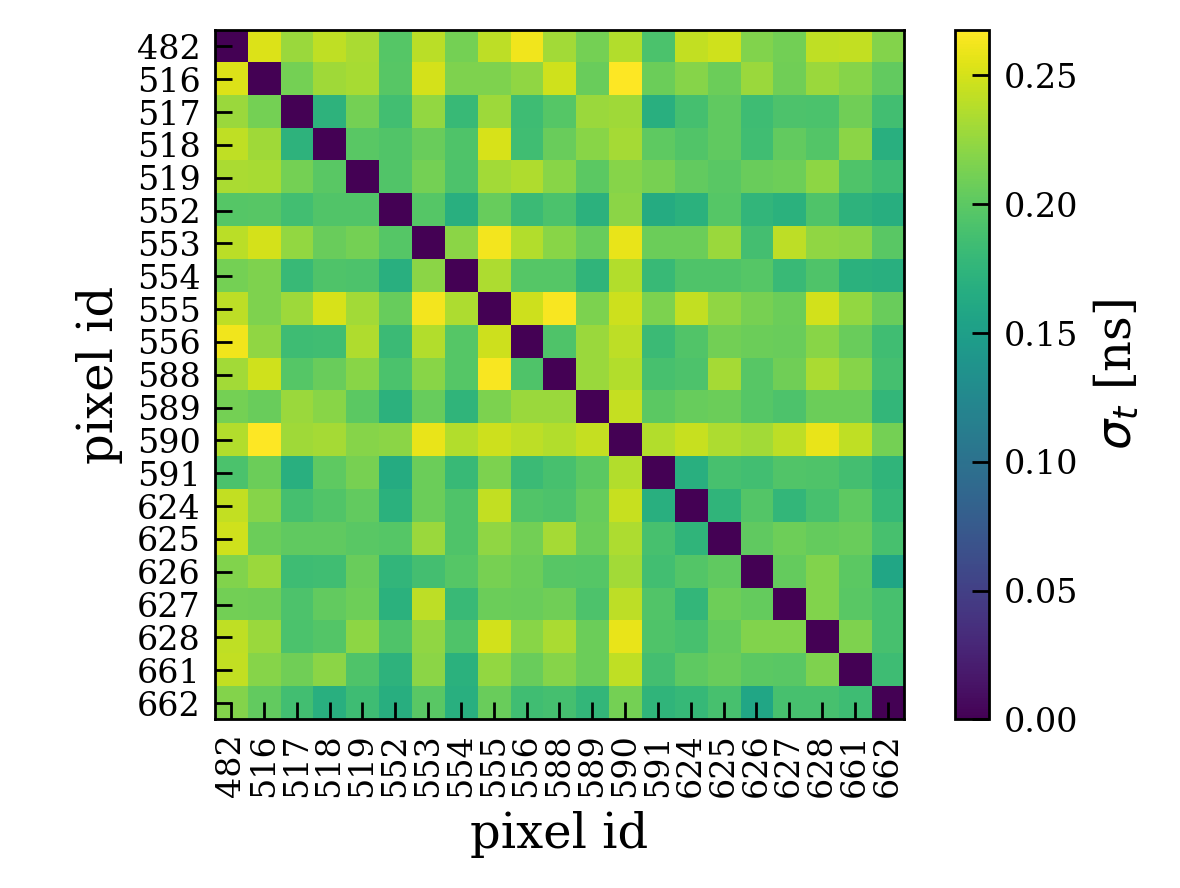}
  \caption{ Measured relative time resolution in 21 neighboring pixels.}
 \label{fig:time}
\end{minipage}
\end{figure}

\item[Temperature mapping:]

The stabilization of the sensor's operational point as a function of temperature is a key feature of the camera design. The cooling system of the PDP and Digicam is described in \cite{camera2016}. In particular, the compensation loop in the slow control system aims at controlling the fluctuations of the operational point of the sensors due to temperature variations.  
The measurement of the temperature over the full camera is presented in figure \ref{fig:mpegain}.  The maximum temperature variation is found to be less than 4$^\circ$~C.

\item[Time resolution:] The timing resolution is yet another important parameter in the camera commissioning to assess the synchronization between different pixels and possible delays in the signal paths issued from each pixel individually. This has been tested by flashing simultaneously a set of 21 neighboring pixels (defined as a cluster) with the same amount of light from CTS, and reconstructing the photon arrival time using template pulse fitting. The maximum time delay between a pair of pixels is +/- 40~ps, which reflects a negligible delay in light injection and detection. The time jitter, i.e. the relative time resolution can reach 0.25~ns within the same cluster, as shown in figure \ref{fig:time}. This translates into 176~ps absolute time resolution\footnote{Considering pixel pairs, the absolute time resolution is obtained by dividing the relative time resolution by $\sqrt{2}$}. This measurement will be extended to the full camera using a unique light source to assess the timing resolution uniformity.     
\end{description}
\subsection{Trigger commissioning}
 
\begin{figure}
\center{
\includegraphics[width = 0.8\textwidth]{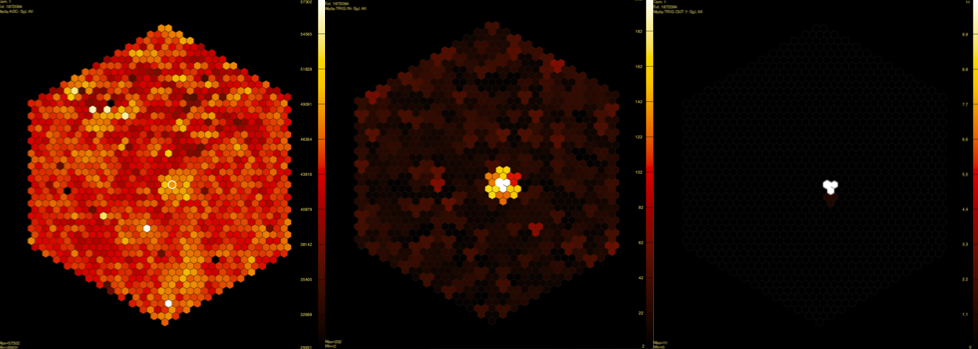}
\caption {One event seen from three different points of view in the camera: (a) left: the 12-bits waveform data used for charge extraction and image reconstruction. (b) middle: the trigger input trace showing higher signal in the cluster where the signal was injected. (c) right: the trigger output trace showing only the central patch of the triggering cluster.}
\label{fig:pixpatch}
}

\end{figure}

 The verification of the proper routing of the signal is crucial for the trigger decision as well as the image reconstruction. It allows to correct for all possible mistakes in the pixel routing on the printed circuit boards, cables, pixel mapping in DigiCam and software decoding of the data stream. In figure \ref{fig:pixpatch} (a), a camera view representing ADC counts in each pixel is shown. Dark pixels represent unsynchronized channels in the digitizer boards at the time of the acquisition. 
A 12-bit waveform of variable length (from 1 to 92 samples of 4~ns) is obtained for every pixel. The baseline is estimated continuously from the average of 1024 samples and subtracted pixel per pixel before being compared to the trigger threshold.  

For each patch, the sum of baseline-subtracted waveforms of each of the 3 pixels is then clipped to 255 ADC (8-bit resolution). The sum can in case  be rescaled prior to the clipping to change the dynamic range of the trigger scale. For the commissioning, we applied a scaling divisor of 4, which will be set to 1 for further data taking. Relying on the validation of the mapping of the pixels in software and hardware, a software bug in the trigger builder could be easily spotted and corrected. 

A cluster is composed of a predefined number of patches around the central patch.  Different combinations of patches defining a cluster (1 patch, 7 patches, 19 patches) have been tested by Monte-Carlo simulations at different conditions of NSB, with the condition that the ADC sum in the cluster passes a certain threshold \cite{camera2016}. From the examined simulations in dark night conditions ($\sim$ 40 MHz/pixel), the lowest energy threshold for detecting gamma rays is achieved with the 7 patches cluster. 
One additional debugging feature of DigiCam is to provide for each triggered event the so-called input and output trigger trace. The input trigger trace is the 8-bit resolution waveform available for each patch which is then summed and compared to the trigger threshold. The output trigger trace is a 1-bit resolution waveform representing the outcome of the trigger decision every 4~ns sample. Both traces have the same duration of the readout window.
An example of a triggered cluster around the central pixel patch is shown in figure~\ref{fig:pixpatch} (b). Its triggered central patch, to which the associated 1-bit trigger output trace is shown in figure \ref{fig:pixpatch} (c).   
  
\section{Camera performance}

\subsection{Trigger rate and readout}

\begin{figure}
\center{
\includegraphics[width = 0.5\textwidth]{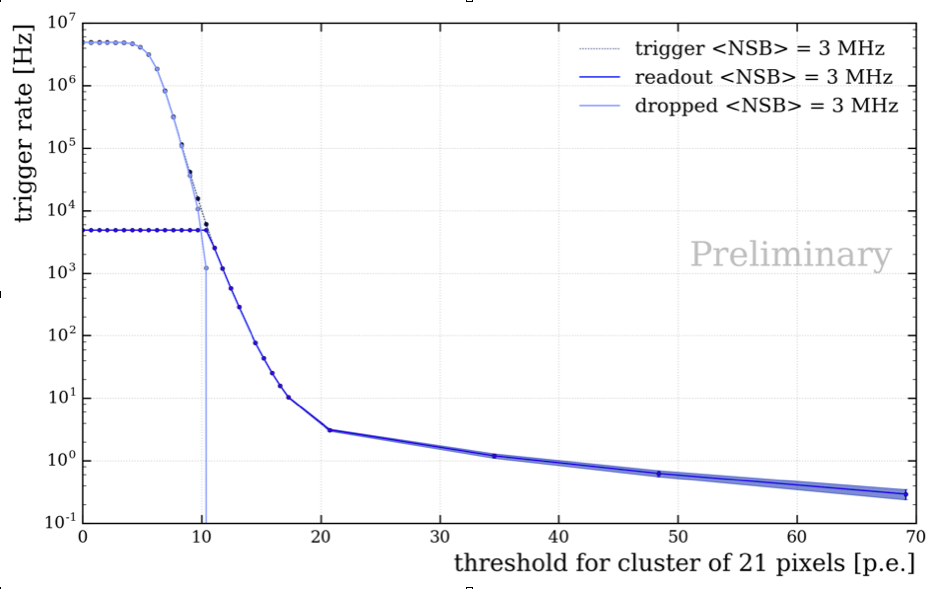}
}

\caption {Readout and trigger rates in dark (night-sky background $\sim$ 3 MHz) derived from Digicam trigger counter (dashed black line). Trigger rate limited by the Digicam readout (dark blue), and the dropped events rate (light blue). }
\label{fig:readout}
\end{figure}

In order to derive the readout efficiency, i.e the fraction of triggered events potentially dropped because of the readout limitation, counters providing the number of events at any time have been implemented in DigiCam.
Trigger rates in dark conditions are shown in figure \ref{fig:readout}. 
As expected, the saturation at low threshold at 5~MHz is due to  the readout window of 50 samples (200~ns) in which only one trigger is registered. This demonstrates that the DigiCam provides dead-time free trigger capabilities in such conditions. 

The readout rate is also shown in figure \ref{fig:readout}, it is saturating here at 4.9~kHz due to the bandwidth limitation. However, DigiCam has a readout capability equivalent to $\sim$435 consecutive events of 50 samples length readout window, meaning that even with randomly distributed triggers generated at the central required frequency of 600~Hz, it would still recover all of them. The full readout chain dead-time will nevertheless be assessed with the official camera server reducing the event rate down to the CTA target of 600 Hz.   
 
\subsection{Bias curves as a function of night-sky background}

In order to validate the triggering over the full camera in different data taking conditions (NSB levels), DC-mode LEDs from the CTS are used to inject light in a single cluster, while the trigger rate is measured after variation of the ADC threshold over the cluster.  

\begin{figure}  
\centering
\begin{minipage}{0.4\textwidth}	
  \includegraphics[width= \textwidth]{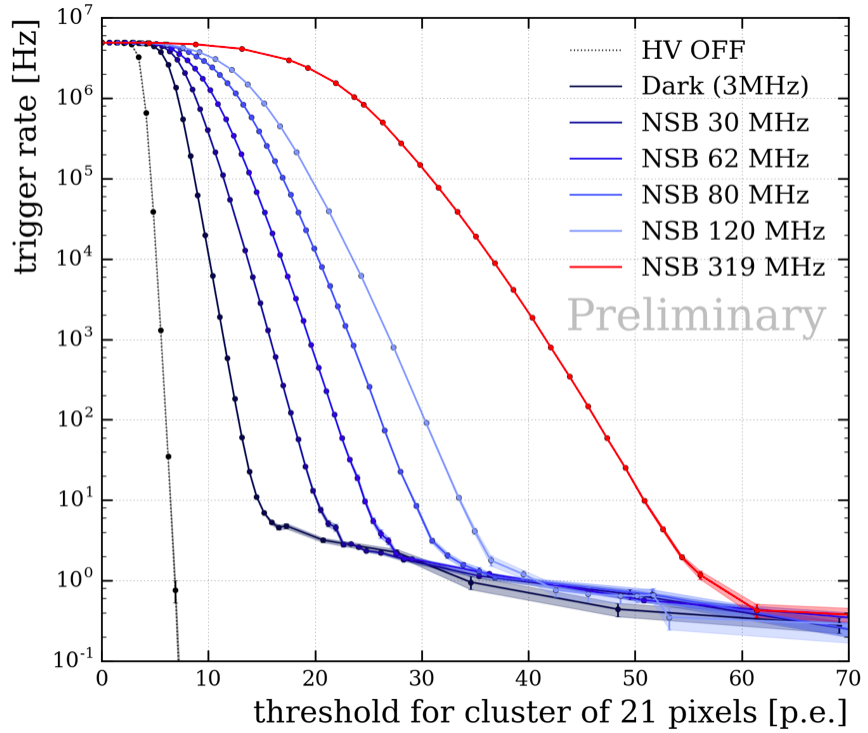}
 \caption{ Trigger rates for different night-sky background levels derived from Digicam trigger counter.}
 \label{fig:trigger_night-sky background}
\end{minipage}
\hspace{1cm}
\begin{minipage}{.43\textwidth}
  \includegraphics[width= \textwidth]{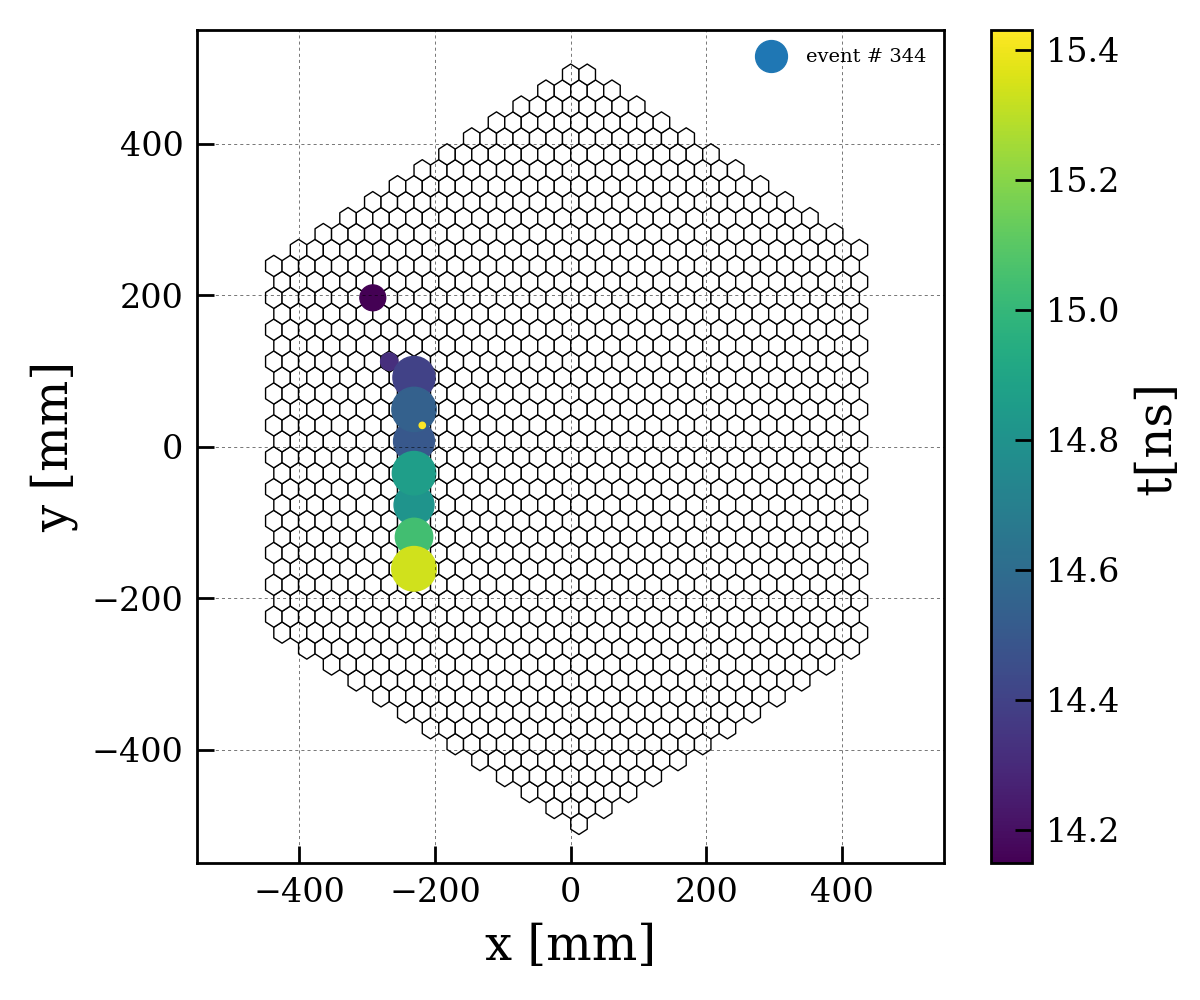}
  \caption{ Reconstructed muon track on the camera in vertical position.}
 \label{fig:vertical}
\end{minipage}
\end{figure}

The result using Digicam trigger counters is shown in figure \ref{fig:trigger_night-sky background} for different NSB levels as a function of the threshold on the total number of PE in the cluster. The trigger rate due to electronic noise only (HV off) reduces drastically above 0.3 PE per pixel on average. At higher NSB conditions, the threshold to apply in order to keep a stable rate is higher. 
Independently from any NSB condition, trigger rates expand to few Hz, despite the higher thresholds applied. After verification of  the charge distributions of the pixels participating in the trigger, we excluded any contribution from events originating in hot pixels. Additional checks rotating the camera vertically led to the identification of cosmic muon tracks crossing the camera. One such event is shown in figure \ref{fig:vertical}. The hit times are reconstructed from the fitted pulse template, and the circle size is proportional to the amplitude of the deposited charge in the camera. With a crossed distance of 30~cm in $\sim$ 1 ns, the reconstructed velocity is compatible with the speed of light as expected.

\subsection {Performance comparison with simulations}

Two independent simulations were used to validate the data acquired so far during the commissioning. A toy Monte-Carlo, presented in \cite{cyrilICRC2017} and CARE\cite{CARE}. In both simulations, pixel calibration parameters presented above were used as input. More particularly, the trigger methodology as well as SiPM properties and dynamic behavior are taken into account. 

\subsubsection*{ADC counts distribution}

\begin{figure}  
\centering
\begin{minipage}{0.45\textwidth}	
  \includegraphics[width= \textwidth]{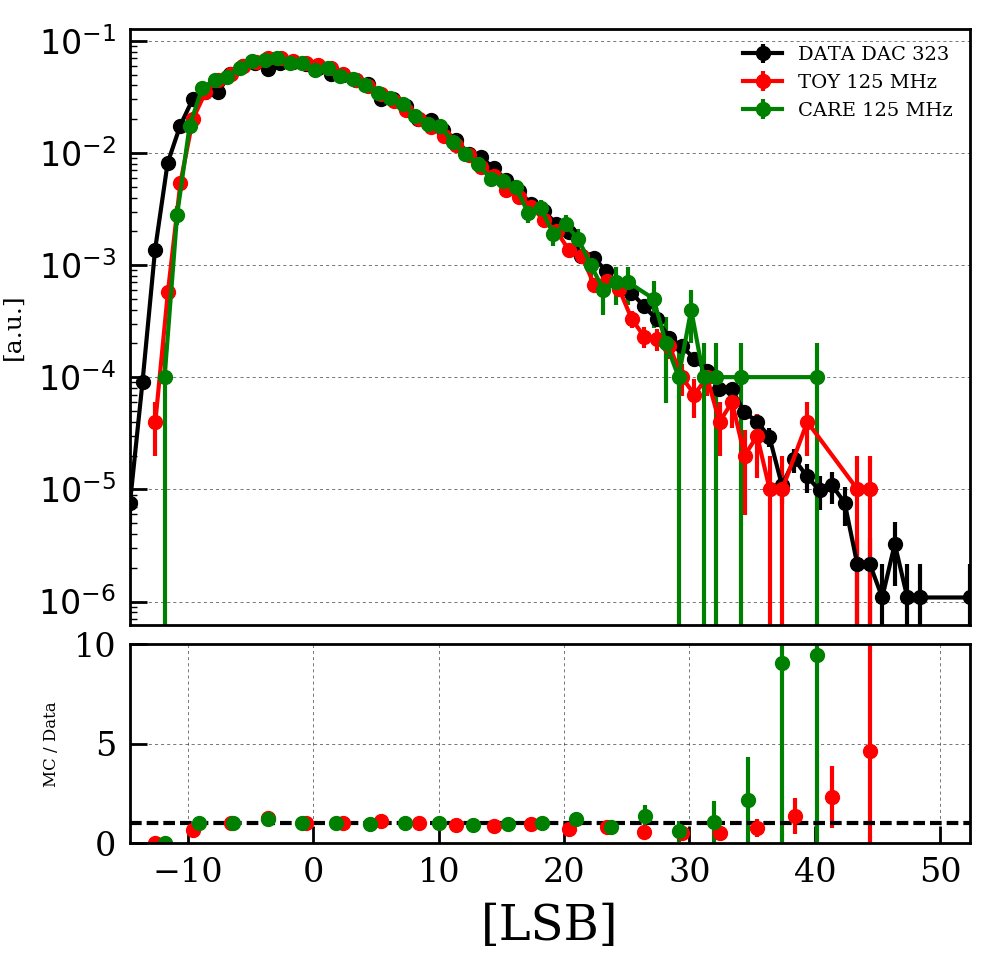}
\end{minipage}
\hspace{1cm}
\begin{minipage}{.45\textwidth}
  \includegraphics[width= \textwidth]{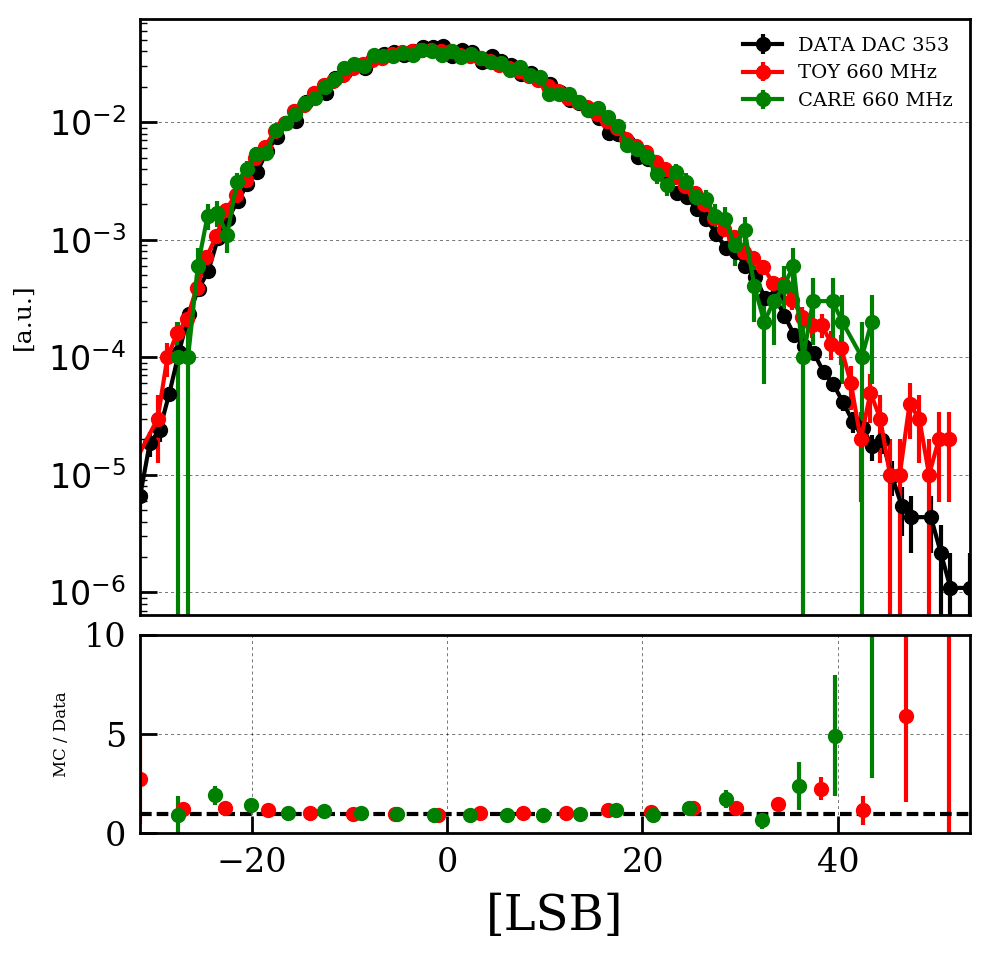}

\end{minipage}
  \caption{ Normalized measured and simulated ADC distributions at 125 MHz/pixel (left) and 660 MHz/pixel (right) night-sky background levels. The dashed black line in the bottom plot represents a MC to data ratio of 1.}
 \label{fig:ADC_dis}
\end{figure}

Low-level comparisons between data and simulations are performed using ADC distributions in different NSB conditions emulated by injected by DC-mode LEDs of the CTS. We show in figure \ref{fig:ADC_dis} a comparison of ADC counts distributions of the digitized signals over the camera for NSB levels of 125 MHz/pixel and 660 MHz/pixel.
The agreement of data and MC is shown in the bottom plot where the ratio of data to MC is plotted, the quoted errors are only statistical. A good agreement is found between both simulations while being completely independent and most importantly reproduce well the data over a large range of data taking conditions. This test has shown the same performances considering dark counts only ($\sim$ 3 MHz/pixel), and at other NSB levels ($\sim$ 40 MHz/pixel and $\sim$ 80 MHz/pixel ).  

\subsubsection*{Trigger rates and trigger efficiencies}

\begin{figure}  
\centering
\begin{minipage}{0.45\textwidth}	
  \includegraphics[width= \textwidth]{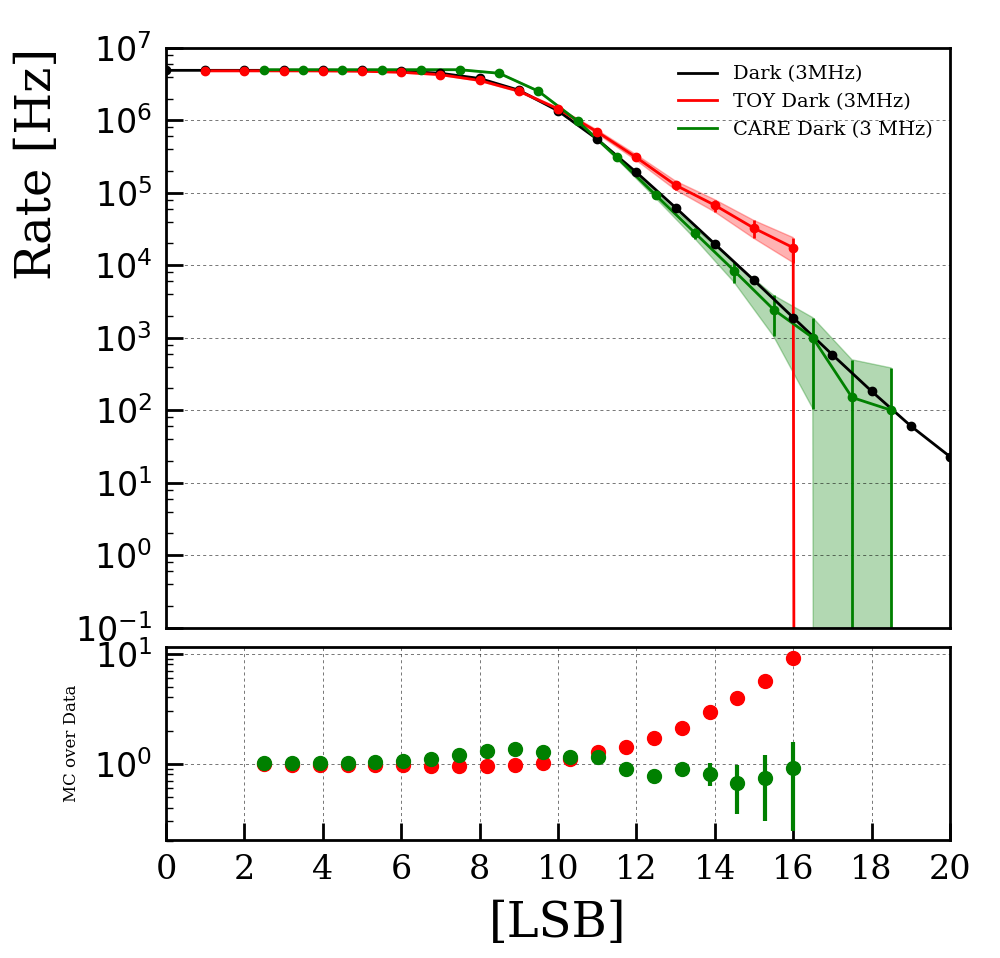}
 \caption{ Bias curves obtained in dark conditions in data and in simulations.}
 \label{fig:mc_vs_data_trigger_rate}
\end{minipage}
\hspace{0.2cm}
\begin{minipage}{.468\textwidth}
  \includegraphics[width= \textwidth]{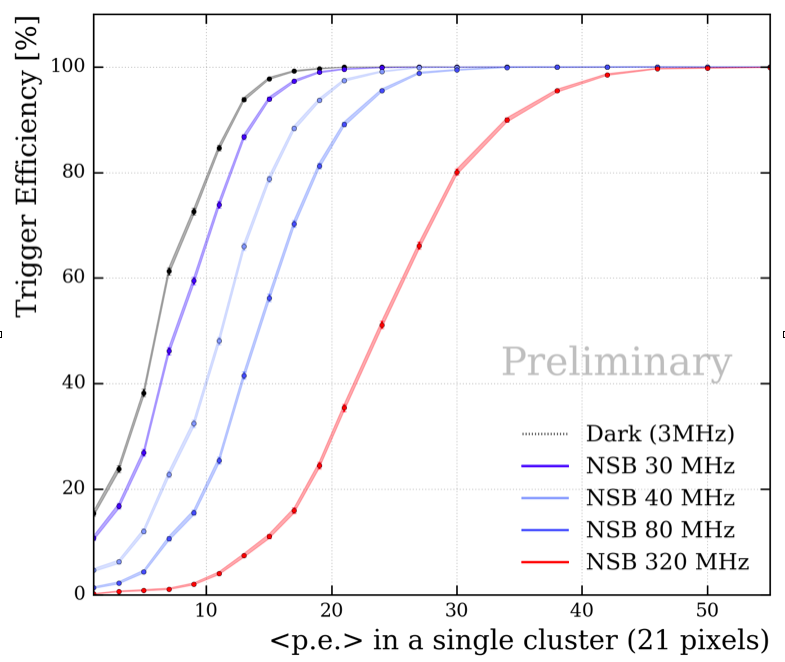}
  \caption{ Trigger efficiency from a cluster in data.}
 \label{fig:trig_effic}
\end{minipage}
\end{figure}

The simulation steps are common to both pipelines; it consists of the SiPM modeling, the signal digitization, the triggering and finally the signal readout.  
The measured trigger rates in the commissioning data as well as the simulations are shown in figure \ref{fig:mc_vs_data_trigger_rate}. The "plateau" is well reproduced by both simulations while the rate evolution as a function of the threshold has a reasonable agreement with both MCs and the data, as shown in the MC to data ratio in the bottom figure. Reproducing the measured bias curves by simulations validates all the simulation steps up to the trigger level. Further cross-checks between data and simulations are being performed at different NSB levels. 
 
The bias curves in figure \ref{fig:trigger_night-sky background} are used to extract the PE level at which the trigger rates reach 500~Hz. A scan in light intensity injected at the central patch of a cluster (equivalent to injecting the same amount of light spread over the full cluster (21 pixels)) is performed. Trigger efficiencies for different NSB levels are shown in figure \ref{fig:trig_effic}. In dark night conditions ($\sim 40$ MHz/pixel), full efficiency is reached at 30 PE in the cluster representing an average of 1.4 PE/pixel. Comparing to simulations with the same trigger topology, 1.7 PE/pixel are found at the same threshold, which is compatible with the result found with data at this NSB level. The trigger configuration consisting of 7 adjacent patches is found to be the most stable (3 PE/pixel on average) with respect to conditions ranging from dark night to half-moon night (660~MHz/pixel). This is also compatible with the above measurements up to 320 MHz, as the maximum variation is below 2.14 PE/pixel.   
  
\section*{Conclusion}

The performance of the SST-1M camera is shown to fully comply with the CTA requirements and design specifications. First data taking with the complete camera using a dedicated test setup was presented and shown to be under control. Upon final validation of the camera simulations, the measured camera response to simulated gamma-ray showers emulated by the CTS will allow for accurate estimations of the signal trigger efficiency considering the real camera dynamic behavior. 

 \section*{Acknowledgements}
 \sloppy
This work was conducted in the context of the CTA Consortium SST-1M Project. 
We gratefully acknowledge support from the University of Geneva, the Swiss National Foundation, the Ernest Boninchi Foundation and the agencies and organizations listed here: 
\url{http://www.cta-observatory.org/consortium_acknowledgments}. In particular we are grateful for support from the NCN grant DEC-2011/01/M/ST9/01891 and the MNiSW grant498/1/FNiTP/FNiTP/2010 in Poland. The authors gratefully acknowledge the support by the projects LE13012 and LG14019 of the Ministry of Education, Youth and Sports of the Czech Republic. This paper has gone through internal review by the CTA Consortium.


\begin{thebibliography}{99}
\bibitem{CTAScience} CTA Consortium, Astropart.Phys. 43 (2013) 3-18
\bibitem{camera2016} M. Heller et al, Eur. Phys. J. C (2017) 77-47
\bibitem{cyrilICRC2017} C. Alispach for the SST-1M project, \pos{PoS(ICRC2017)800} (these proceedings)
\bibitem{CARE} $http://otte.gatech.edu/care/tutorial/$
\bibitem{matthieuSPIE} J. A. Aguilar et al., Proc. SPIE 9915, High Energy, Optical, and Infrared Detectors for Astronomy VII, 99152T (27 July 2016)

\end{thebibliography}
\end{document}